\documentclass[12pt]{article}
\begin{document}
\title{Mass matrices for quarks and leptons in triangular form}
\author{ Stanis\l aw Tatur
\\ Nicolaus Copernicus Astronomical Center,\\ Polish Academy of
Sciences,\\ Bartycka 18, 00-716 Warsaw, Poland. \\ \\ \and Jan
Bartelski\\ Institute of Theoretical Physics,\\ Faculty of Physics,
Warsaw University,\\ Ho$\dot{z}$a 69, 00-681 Warsaw, Poland. \\ }
\date{}
\maketitle

\begin{abstract}
\noindent We assume that all quark and lepton  mass matrices have
upper triangular form. Using all available experimental data on
quark and lepton masses and mixing angles we make a fit in which we
determine mass matrices elements. There are too many free parameters
and our solutions are not unique. We look for solutions with small
non diagonal mixing matrix elements.  In order to reduce the number
of free parameters we assume that the matrix element $ (M)_{13}$
vanishes in \underline{all} mass matrices. Such universal assumption
was drown from considering different numerical solutions. The lepton
sector, due to large mixing angles and very small errors for charged
lepton masses, is more restrictive then quark sector. We present the
solution in this case. The absolute values of neutrino masses are
not fixed. The another possibility of reducing number of free
parameters was considered by us before. With the additional
assumption motivated by $SU(5)$ symmetry which connects mixing in
right handed down quarks with left handed charged leptons we get a
solution in which observed Cabibbo-Kobayashi-Maskawa mixing for
quarks comes mainly from non diagonal terms in up quark mass matrix.
\end{abstract}

\newpage
The understanding of masses and mixing for quarks and leptons, after
discovery that neutrinos are massive \cite{osc}, is one of the most
interesting and hot problems \cite{rev} in the standard model. In
this paper we shall consider rather unusual idea of mass matrices in
triangular form. Such idea was proposed  in \cite{jbst}. In that
paper, taking into account very specific additional assumptions,
mass matrices for up and down quarks, neutrinos and charged leptons
were given together with mixing matrices for left and right handed
components of fields. We considered model which neglects CP
violation as well as the one that took it into account in quark
sector. Neutrinos, contrary to now popular assumption of being
Majorana particles, were treated as Dirac particles.

In this paper we also assume that the neutrino is a Dirac particle.
We believe that because there is no positive result of double
$\beta$ decay experiment it is still a viable hypothesis. It could
be worthwhile (smaller number of parameters) to study consequences
of this assumption. We have shown in \cite{jbst}, first considering
real matrices, that inclusion of CP violation for quarks was not a
problem. In this paper we will consider real matrices neglecting CP
violation because we believe after experiences from \cite{jbst} that
it will be not difficult to include it. In addition $(M)_{13}$
element of lepton mixing matrix seems to be very small (if at all
different from zero) and we do not have any information about CP
violating phase.

How we can justify the assumption of using mass matrices in
triangular form. We believe that left handed $SU(2)$ doublets and
not right handed singlets are responsible for mixing. After
spontaneous symmetry breaking our basic objects that we have in
lagrangian are mass matrices for up and down quarks, neutrinos and
charged leptons. When we make a decomposition of mass matrix into
diagonal matrix and two unitary matrices, taking into account that
we have strongly hierarchical mass spectrum for up and down quarks
as well for charged leptons and not so strongly  for neutrinos, we
will find out that matrix elements above diagonal influence mostly
mixing for left handed components and below diagonal right handed
components. If we want to have mixing determined by left handed
components (active in interactions) we assume that in our  mass
matrices  matrix elements responsible for mixing of right handed
(below diagonal) components vanish. In this way we end up with upper
triangular mass matrices that (with hierarchical mass spectrum) give
very asymmetrical mixing much stronger for left handed then for
right handed components. In this paper we continue to use the
assumption of triangular form for quark and lepton mass matrices but
we do not take into account all other specific assumptions we made
in \cite{jbst}. The masses of quarks and leptons and the
corresponding errors are very different spanning many orders of
magnitude. To cope with this and the fact that for neutrinos  we do
not know masses but only mass squared differences we will use, as in
\cite{jbst} $\chi^2$ function (the least square method). Knowing
measured mixing matrices for left components and their errors and
masses of quarks and leptons with corresponding errors we will try
to calculate mass matrices for quarks and leptons assumed in upper
triangular form.

In general we have too many free parameters and we can not find
solution that is unique. We give examples of solutions with
relatively small non diagonal matrix elements in mass matrices. The
question is can we reduce the number of free parameters by assuming
that some of them vanish. It is not enough to calculate formally
number of free parameters. In the case of neutrinos and charged
leptons it is not possible to get solution (with small $\chi^2$
value) when neutrino mass matrix is diagonal and whole mixing is in
charged lepton sector in spite of the fact that formally we have
enough free parameters to get a solution. We can get analogous
solution for up and down quarks. We discuss in detail the different
cases for quarks and leptons when we have enough free parameters and
could not find the solution with small $\chi^2$ value. It seems that
lepton sector is more restrictive then quark sector. The mixing for
leptons is much stronger then for quarks and charged lepton masses
are known with high accuracy (specially electron and muon masses) in
comparison with up and down quark masses. As a result of our
numerical analysis we have found out that when in addition to the
assumption of upper triangular form of  mass matrices we assume that
in neutrino and charged leptons mass matrix matrix element
$(M)_{13}$ vanishes we can find a solution. This assumption is
generalized and is made for all multiplets of up and down quarks
neutrinos and charged leptons so they are all treated in the same
way. Such assumption can be justified by the fact that measured
mixing matrices for quarks and leptons have matrix element
$(U)_{13}$ very small (smaller then the others, in the case of
neutrinos it is not well known but very close to zero or even zero)
so we expect that mixing of first and third generation comes as a
result of mixing of other generations. Absolute values of neutrino
masses can not be fixed in this way.

At the end we  will give rather exotic example of solution in which
number of free parameters is reduced by 3 by a relation between
different mixing matrices. We thought at the beginning that it could
be an alternative to the assumption of vanishing mass matrix
elements. We will show that it is possible to get solutions for
triangle mass matrices when we assume connection between left and
right handed mixing matrices. It is suggested by $SU(5)$ symmetry
condition that right handed components of down quarks transform in
the same way as left handed components of charged leptons (belonging
to the same $SU(5)$ multiplet). The errors for electron and muon
masses are very small so the mixing of charged leptons is rather
small and that has a consequence that the contribution of mixing to
CKM matrix from down quarks is not big and comes mostly from the
mixing of up quarks. With our additional assumption for all mass
matrices $(M)_{13}=0$ we can not find such solution.

Let us  introduce the notation for quarks. We will follow the
notation used in [1] and repeat introductory formulas. The
$SU(3)*SU(2)*U(1)$ gauge invariant Yukawa interactions for quarks
are given by
\begin{equation}
-L_{Y}=\bar{Q}_{i} (Y_{u})_{ij} u_{Rj} H^{\dagger}+
\bar{Q}_{i}(Y_{d})_{ij}d_{Rj}H + h.c. ,
\end{equation}

\noindent where $Q_{i}$ denote the $SU(2)$ doublets of left-handed
quarks and $u_{R}$, $d_{R}$ are the right-handed up and down-type
quarks respectively. The Yukawa couplings $Y_{u}$ and $Y_{d}$ are
$3\times3$ matrices (i,j are the generation indices) and $H$ is
the $SU(2)$ doublet Higgs field. After the electroweak symmetry
breaking, these Yukawa interactions lead to the following quark
mass terms
\begin{eqnarray}
-L_{m}&=&\bar{u}_{Li}(M_{u})_{ij}u_{Rj}+ \bar{d}_{Li}(M_{d})_{ij}d
_{Rj}+ h.c., \\ (M_{u})_{ij}&=&(Y_{u})_{ij}v, \nonumber \\
(M_{d})_{ij}&=&(Y_{d})_{ij}v, \nonumber
\end{eqnarray}

\noindent where $v$ is vacuum expectation value of the neutral
component of the Higgs field $H$.

Mass matrices $M_{u}$ and $M_{d}$ (considered by us as basic
objects) can be diagonalized using two unitary matrices $U$ and $V$
\begin{eqnarray}
M_{u}&=&U_{u}M_{u}^{D}V_{u}^{\dagger}, \\
M_{d}&=&U_{d}M_{d}^{D}V_{d}^{\dagger}.
\end{eqnarray}

The diagonal matrix elements (of $M_{u}^{D}$ and $M_{d}^{D}$)
correspond to the experimentally observed mass eigenvalues. The
matrices $U$ and $V$ describe mixing between states with definite
flavor $u$, $d$ and with definite mass $u'$, $d'$ for left handed
and right handed components
\begin{eqnarray}
u_{L}&=&U_{u}u'_{L}, \nonumber \\  u_{R}&=&V_{u}u'_{R},  \\
d_{L}&=&U_{d}d'_{L}, \nonumber \\  d_{R}&=&V_{d}d'_{R}. \nonumber
\end{eqnarray}

The generation mixing in the charged weak current after expressing
in terms of fields with definite mass is  described by the
Cabibbo-Kobayashi-Maskawa (CKM) matrix \cite{CKM} which consists
of two unitary matrices
\begin{equation}
U_{CKM}=U_{u}^{\dagger}U_{d}.
\end{equation}

 In the similar way, assuming that
neutrinos are traditional Dirac particles, we have mixing for the
charged weak current in the lepton sector described by
Pontecorvo-Maki-Nakagawa-Sakata (PMNS) matrix \cite{MNS}

\begin{equation}
U_{MNS}=U_{l}^{\dagger}U_{\nu}.
\end{equation}

\noindent The relations for neutrino mass $M_{\nu}$ and charged
leptons $M_{l}$ are:

\begin{eqnarray}
M_{\nu}&=&U_{\nu}M_{\nu}^{D}V_{\nu}^{\dagger}, \\
M_{l}&=&U_{l}M_{l}^{D}V_{l}^{\dagger}.
\end{eqnarray}

\noindent As before diagonal matrix elements of $M_{\nu}^{D}$ and
$M_{l}^{D}$ correspond to the experimentally observed mass
eigenvalues.

We have got experimental data for CKM and PMNS matrices. We will use
for CKM matrix and the errors of matrix elements values given in
\cite{hock}. The values and the errors for up and down quark masses
are taken from the \cite{pdg}. The parameters of PMNS matrix and
corresponding errors are taken from recent updating of the fit
\cite{val} taking into account recent results from KamLAND and MINOS
experiments given in \cite{min}. Mass squared differences of
neutrino masses are taken from the same fit \cite{val} and masses of
charged leptons and their errors are taken from \cite{pdg}. The
masses of charged leptons are known with extremely high accuracy in
comparison with quark and neutrino ones. For neutrinos only mass
squared differences are known. That was the reason why in
\cite{jbst}, in order to fit mass matrices for quarks and leptons in
triangular form to experimental data, we used numerical
minimalization of $\chi^2$ function  taking into account
experimental errors. The corresponding modified function was used
for PMNS matrix and mass values for neutrinos and charged leptons.
Minimalization of $\chi^2$ gives most probable value of parameters
or could give numerical solution to equations (as it was used in
\cite{jbst}. When there are too many free parameters we can get one
of many possible solutions. In \cite{jbst} we have also assumed that
we have upper or lower triangular matrices for up, down, neutrinos
and charged leptons and that matrices diagonalizing right handed
components are the same in the weak isotpic spin multiplets.

In the present paper we relax these additional assumptions. The left
handed components of  all quarks and leptons take part in charged
weak current interactions. We want to have a situation that mixing
in active left handed components determines the whole mixing. We
assume that mass matrices for up, down quarks, neutrinos and charged
leptons are upper triangular matrices. When we have upper triangular
mass matrices with hierarchical diagonal elements decomposition into
two unitary matrices and diagonal matrix we have very simple
situation. Mixing matrix elements for left handed components $U$ are
determined to the first order by the ratios of non diagonal and
corresponding diagonal mass matrix elements ($(U)_{12} \simeq
(M)_{12}/(M)_{22}$, ($(U)_{23} \simeq (M)_{23}/(M)_{33}$, ($(U)_{13}
\simeq (M)_{13}/(M)_{33}$) and matrix elements for right handed
components $V$ are determined by left handed mixing elements
multiplied by corresponding ratios of diagonal masses ($
(M)_{11}/(M)_{22}$, $ (M)_{11}/(M)_{33}$, $ (M)_{22}/(M)_{33}$).
Ortogonality of matrices $U$ and $V$ must of course be taken into
account. When we have hierarchical masses the mixing in $V$ is
determined by $U$ and much smaller. All that will clearly be seen in
our example of the first solutions. In \cite{jbst} we used
artificial errors for electron and muon masses. We want to remind
that masses of electron and muon are known with extremely high
accuracy in comparison with all other particles. It seems that error
in the mass of electron is comparable with the expected naively mass
of heaviest neutrino. One can expect that these small errors can
give some limitations on mixing in charged mass matrices because
when we diagonalize them you have to end up with the values within
small errors. We will try to see what are the consequences of small
errors in masses of charged leptons in comparison with quarks.

Following \cite{jbst} for quarks we will  minimalize the function

\begin{eqnarray}
\chi^2&=&\sum_{ij}
\frac{((U_{u}^{\dagger}U_{d})_{ij}-U_{CKMij}^{exp})^2}{(\Delta
U_{CKMij}^{exp})^2} + \sum_{i}
\frac{(m_{di}^{D}-m_{di}^{exp})^{2}}{(\Delta m_{di}^{exp})^{2}}+
\nonumber \\ &+& \sum_{i}
\frac{(m_{ui}^{D}-m_{ui}^{exp})^{2}}{(\Delta m_{ui}^{exp})^{2}}.
\end{eqnarray}

There are 6 free parameters (3 diagonal and 3 non diagonal) for
upper triangular matrix $M_{d}$ and 6 parameters for upper
triangular matrix $M_{u}$ (in principle we could also use lower
triangular matrix for $M_{u}$ like in \cite{jbst}). As we mentioned
before the values $m_{di}^{exp}=(m_{d},m_{s},m_{b})$,
$m_{ui}^{exp}=(m_{u},m_{c},m_{t})$ and corresponding errors $\Delta
m_{di}^{exp}$, $\Delta m_{ui}^{exp}$ are taken from the Particle
Data Group  \cite{pdg} and $U_{CKM}^{exp}$ and corresponding errors
from \cite{hock}. The values of $U_{CKM}^{exp}$ are expressed in
terms of angles to have unitarity satisfied to high degree and then
we use calculated in this way matrix elements. For neutrinos and
charged leptons we use the same procedure with obvious
modifications. In the $\chi^2$ function we fit $U_{CKM}$ or
$U_{PMNS}$ matrix elements and masses of quarks or leptons using 12
free parameters ( 2 mass matrices, 3 diagonal and 3 non diagonal
elements). There are more parameters we want to determine then
independent fitted quantities so we can not uniquely determine them.
Anyhow, we do not understand the mass scales and big mass ratios for
different generations of quarks. So it could happen that just
because of big difference in mass scales non diagonal matrix
elements are small in comparison with diagonal ones (for quantum
systems very small mixing between very different energy levels)
giving rather small mixing in $U_{CKM}$ matrix. We will look for
example  of this type of solution.

 We will start with $U_{CKM}$ matrix and up and down quark
mass matrices and we consider the fit with relatively small
parameters (elements of mass matrices are given in $MeV$)

\begin{equation}
M_{d} =\left( \begin{array}{ccc} 5.11523&20.03&10.5861
\\0&92.9391&172.911 \\0&0&4196.42
\end{array} \right)
\end{equation}

\begin{equation}
U_{d} =\left( \begin{array}{ccc} 0.97741&0.21134&0.00252
\\-0.21126&0.97656&0.04119 \\0.00624&-0.04079&0.99915
\end{array} \right)
\end{equation}

\begin{equation}
M_{d}^{D}=\left( \begin{array}{ccc}5&0&0
\\0&95&0 \\0&0&4200
\end{array} \right)
\end{equation}

\begin{equation}
V_{d}^{\dagger}=\left( \begin{array}{ccc}
0.99993&-0.01138&7.43\times 10^{-6}
\\0.01138&0.99993&-0.00092 \\3.08
\times 10^{-6}&0.00924&1
\end{array} \right)
\end{equation}

and

\begin{equation}
M_{u} =\left( \begin{array}{ccc} 2.2503&-20.37&-105.96
\\0&1249.83&111.654 \\0&0&172500
\end{array} \right)
\end{equation}

\begin{equation}
U_{u} =\left( \begin{array}{ccc} 0.99987&-0.01630&-0.00061
\\0.01630&0.99987&-0.00064 \\0.00062&0.00064&1
\end{array} \right)
\end{equation}

\begin{equation}
M_{u}^{D}=\left( \begin{array}{ccc}2.25&0
\\0&1250&0 \\0&0&172500
\end{array} \right)
\end{equation}

\begin{equation}
V_{u}^{\dagger}=\left( \begin{array}{ccc} 1&0.00003&8.15 \times
10^{-9}
\\-0.00003&1&4.62
\times 10^{-6} \\-8.01 \times 10^{-9}&-4.62 \times 10^{-6}&1
\end{array} \right)
\end{equation}

The $\chi^2 $ is practically zero (we get the number of order
$10^{-14}$) so we can treat these parameters as a numerical
solution of Eqs. (3,4) and
 Eq. (6). We see that relations between matrix elements
of  $U$ and $V$ mentioned before and mass matrix elements are
satisfied. The obtained solution is not so different in character
 from that obtained with different additional assumptions in
 \cite{jbst}. One can easily give (because we have 3 free
 parameters) examples with bigger non diagonal matrix elements
 giving the same $U_{CKM}$ mixing matrix.

When we look for small mixing parameters in case of neutrinos and
charged lepton mass matrices we get (in $eV$)

\begin{equation}
M_{\nu} =\left( \begin{array}{ccc} 0.00265&0.00529&0.00327
\\0&0.01050&0.03329 \\0&0&0.03502
\end{array} \right)
\end{equation}

\begin{equation}
U_{\nu} =\left( \begin{array}{ccc} 0.82188&0.56353&0.083388
\\-0.44863&0.55008&0.70437 \\0.35107&-0.61632&0.70491
\end{array} \right)
\end{equation}

\begin{equation}
M_{\nu}^{D}=\left( \begin{array}{ccc}0.00221&0&0
\\0&0.00899&0 \\0&0&0.04904
\end{array} \right)
\end{equation}

\begin{equation}
V_{\nu}^{\dagger}=\left( \begin{array}{ccc}
0.98611&-0.16461&0.02214
\\0.16603&0.97333&-0.15828 \\0.00451&0.15975&0.98715
\end{array} \right)
\end{equation}

and (in $MeV$)

\begin{equation}
M_{l} =\left( \begin{array}{ccc} 0.510999&-0.03867&-0.07211
\\0&105.658&-0.71477 \\0&0&1776.99
\end{array} \right)
\end{equation}

\begin{equation}
U_{l} =\left( \begin{array}{ccc}
1&-0.00037&-0.00004\\0.00037&1&-0.00040
\\0.00004&0.00040&1
\end{array} \right)
\end{equation}

\begin{equation}
M_{l}^{D}=\left( \begin{array}{ccc}0.510999&0&0
\\0&105.658&0 \\0&0&1776.99
\end{array} \right)
\end{equation}

\begin{equation}
V_{l}^{\dagger}=\left( \begin{array}{ccc} 1&1.77\times
10^{-6}&1.17\times 10^{-8}
\\1.77\times
10^{-6}&1&0.00002 \\-1.17 \times 10^{-8}&-0.00002&1
\end{array} \right)
\end{equation}

These are just examples of solutions with small non diagonal terms
in mass matrices $M_{u}$ and $M_{l}$. We have relatively small
mixing in $U_{u}$ and $V_{u}$ coming from $M_{u}$ and in $U_{l}$ and
$V_{l}$ connected with $M_{l}$. The mixing in $U_{CKM}$ and
$U_{PMNS}$ comes mainly from $U_{d}$ and $U_{\nu}$ connected with
non diagonal terms of matrices $M_{d}$ and $M_{\nu}$. For the quarks
we have 3 free parameters and for leptons 4 (we do not know values
of masses of neutrinos only mass squared differences) so we can find
many very different solutions. In the CKM matrix
$U_{CKM}=U_{u}^{\dagger}U_{d}$ mixing can come from both matrices
(the solution where most of the mixing in $U_{CKM}$ comes from
$U_{u}$ is possible. That it is not the case for leptons. We do not
have solutions where in $U_{PMNS}$ mixing matrix most of the mixing
comes from charged lepton mixing matrix $U_{l}$. The mixing in the
lepton case is much stronger then for quarks and quark masses are
known with much less accuracy then masses of charged leptons.

In the case of leptons situation is a bit different because of very
small errors in masses of $e$ and $\mu$ leptons. The diagonal masses
are nearly equal to the measured charged lepton masses and non
diagonal masses are really tiny. The biggest error is for $\tau$
lepton mass. We have the strongest dependence on only one parameter
namely element $(M_{l})_{23}$ of charged lepton mass matrix. One can
numerically try to find out how big this element could be still
having solutions to our equations. We will give an example of the
solution for leptons where we have relatively strong mixing because
of "big" non diagonal matrix elements in $m_{l}$ mass matrix for
charged leptons.

\begin{equation}
M_{\nu} =\left( \begin{array}{ccc} 0.00971&0.00278&-0.00038
\\0&0.01493&0.02908 \\0&0&0.03942
\end{array} \right)
\end{equation}

\begin{equation} M_{l}=\left(
\begin{array}{ccc}0.51403&-11.3797&-9.29737
\\0&105.647&-190.258 \\0&0&1766.71
\end{array} \right)
\end{equation}

In this case $U_{\nu}$ and $U_{l}$ are given by

\begin{equation}
U_{\nu} =\left( \begin{array}{ccc} 0.86365&0.50407&0.00323
\\-0.39334&0.66988&0.62972 \\0.31527&-0.54513&0.77681
\end{array} \right)
\end{equation}

\begin{equation} U_{l} =\left( \begin{array}{ccc}
0.99411&-0.10826&-0.00519
\\0.10708&0.98843&-0.10744 \\0.01676&0.10625&0.99420
\end{array} \right)
\end{equation}

This is the solution that contrary to solution given in Eqs. (19-26)
corresponding to small non diagonal mass matrix elements gives
relatively big non diagonal matrix elements for $M_{l}$. The mixing
for charged leptons is still rather small in comparison with that
for neutrinos. That makes that situation for leptons is different
from that for quarks. In the case where non diagonal terms in
charged lepton mass are relatively big the mixing in charged lepton
mass matrix $U_{l}$ is small and is only a small correction to
$U_{\nu}$ being mainly responsible for the value of $U_{PMNS}$.
Comparing with Eq. (27) with Eq. (19) we see that there is some
increase in diagonal matrix elements in $M_{\nu}$ and that
corresponds to bigger neutrino masses $m_{1\nu}=9.1\;  meV$,
$m_{2\nu}=12.6\; meV$, $m_{3\nu}=49.8\;  meV$ in comparison with
$m_{1\nu}=2.2\; meV$, $m_{2\nu}=9.0\;  meV$, $m_{3\nu}=49.0\;  meV$
in Eq. (21). Contrary to quarks, in the lepton case we have rather
small range of non diagonal terms in charged lepton mass matrix. The
mixing given by $U_{l}$ and $V_{l}$ is small and the values of
$U_{PMNS}$ are dominated by non diagonal terms in $M_{\nu}$ and
mixing matrix $U_{\nu}$.

Up to now we have considered examples of solutions with too many
free parameters which we want to determine. One can restrict the
freedom of possible solutions by some additional conditions e.g. as
in \cite{jbst} when the equality of $V$ matrices in the same weak
isospin multiplet was assumed. We will present mass matrices when we
limit the number of parameters in mass matrices by putting 3 of them
equal to zero. We will start with extreme cases when when all
vanishing mass matrix elements are in up or down quark mass matrix.
In the case of quarks we get for diagonal $M_{u}$ :

\begin{equation}
M_{d} =\left( \begin{array}{ccc} 5.13392&21.5375&16.0237
\\0&92.6034&175.388 \\0&0&4196.3
\end{array} \right)
\end{equation}

and $U_{d}=U_{CKM}$, whereas for diagonal $M_{d}$ we have:

\begin{equation}
M_{u} =\left( \begin{array}{ccc} 2.31044&-284.247&996.199
\\0&1218.37&-7167.79 \\0&0&172348
\end{array} \right)
\end{equation}

\noindent and of course $U_{u}^{\dagger}=U_{CKM}$.

These are extreme cases when mixing in $U_{CKM}$ matrix either comes
from non diagonal terms in down quark mass matrix or from non
diagonal terms in up quark matrix. We can also numerically study
solutions when three zero matrix elements are distributed among two
such matrices. It is easy to check that in spite of the fact that
formally we have enough free parameters (3 zero matrix elements in
$M_{u}$ and $M_{d}$ matrices) such solutions do not exist
($\chi^{2}$ is not small). When $(M_{d})_{13}=0$, $(M_{u})_{12}=0$,
$(M_{u})_{13}=0$ or we change indexes $u$ and $d$ namely
$(M_{u})_{13}=0$, $(M_{d})_{12}=0$, $(M_{d})_{13}=0$ there is no
solutions ($\chi^{2}>700$ or $\chi^{2}>1400$). There is also no
solution when $(M_{d})_{23}=0$, $(M_{u})_{13}=0$, $(M_{u})_{23}=0$.
That means that when we have number of free parameters equal to the
number of experimental data we not always can find a solution.  On
the other hand when $(M_{d})_{13}=0$, $(M_{u})_{13}=0$ and
$(M_{u})_{23}=0$ or $(M_{d})_{13}=0$, $(M_{u})_{12}=0$,
$(M_{u})_{23}=0$ we can easily find a solution.

In the case of leptons when charged lepton mass matrix is diagonal
(non diagonal elements are put equal to zero) we get:

\begin{equation}
M_{\nu} =\left( \begin{array}{ccc} 0.00385&0.00510&0.00328
\\0&0.01100&0.03323 \\0&0&0.03509
\end{array} \right)
\end{equation}

\noindent and $U_{PMNS}=U_{\nu}$. It is also possible to have
solution for $M_{\nu}$ with $ m_{1\nu}$ close to zero.

We get

\begin{equation}
M_{\nu} =\left( \begin{array}{ccc} 10^{-12}&0.00549&0.00329
\\0&0.01005&0.03337 \\0&0&0.03494
\end{array} \right)
\end{equation}
We want to stress that with diagonal mass matrix for charged leptons
we still have one free parameter (we do not know absolute values of
neutrino masses) and the smallest neutrino mass equal to zero is not
excluded.

In the case of leptons contrary to quarks, where we could have
solution with diagonal $M_{d}$ (Eq. (32)) but it is not possible (at
least with high probability $\chi^{2}>170$) to have a satisfactory
solution that neutrino mass matrix is diagonal and the whole strong
mixing that we have in $U_{PMNS}$ comes from charged leptons. It is
understandable from the discussion of mixing in charged lepton
sector. With relatively small mixing in the charged lepton sector we
can not reproduce very strong mixing observed in $U_{PMNS}$. The
question is can we split three zeros between neutrino mass matrix
and charged lepton mass matrix to have $U_{\nu}$ and $U_{l}$ mixing
matrices that reproduce strong mixing in $U_{PMNS}$. To reproduce
$U_{PMNS}$ with two big mixing angles we need nonzero matrix
elements $(M_{\nu})_{12}$ and $(M_{\nu})_{23}$. Having only these
matrix non diagonal matrix elements different from zero with all the
other non diagonal matrix elements in $M_{l}$ equal to zero we get
$\chi^{2}$ bigger then 2. With $(M_{\nu})_{12}$ and $(M_{\nu})_{23}$
different from zero and  $(M_{l})_{12}$ or $(M_{l})_{13}$ or
$(M_{l})_{23}$ we always get $\chi^{2}$ bigger then 2. With 3
additional zeros in mass matrices we still have 1 free parameter
connected with lack of scale for neutrino masses. Calculating
formally number of parameters we should have a solution. The next
possibility is to add one more non zero non diagonal matrix element.
If we want to treat neutrinos and charged leptons in the same way
then the choice is obvious. We assume that in charged lepton mass
matrix non diagonal matrix element $(M_{l})_{23}$ is different from
zero. In this case we can find a solution. That means that guided by
numerical calculations we assumed that matrix elements
$(M_{\nu})_{13}$ and $(M_{l})_{13}$ are equal to zero. We will also
assume that for mass matrices of up and down quarks. It is the
assumption corresponding for triangular matrices to that considered
and advertised for quarks by Fritzsch, idea of having mixing only
between neighbored flavors \cite{fritz}. On the other hand from
experiment we know that  $(U_{CKM})_{13}$ is small and in
$(U_{PMNS})_{13}$ is small (not well known) and still consistent
with zero so assuming that in basic mass matrices $(M)_{13}$
vanishes is not so unnatural. In some sense the situation is very
simple. We ask is it possible to reduce number of solutions by
assuming that some mass matrix elements are equal to zero. If we
want to add in a universal way (in the same way for up and down
quarks, neutrinos and charged leptons) condition of vanishing mass
matrix elements we do not get a solution with 2 zeros in every mass
matrix. The next possibility is to consider one vanishing mass
matrix element in every mass matrix. Discussion of numerical
solutions in lepton sector has helped us to localize  these
vanishing mass matrix elements. We will present a solution with two
additional zeros when matrix elements $(M_{\nu})_{13}$ and
$(M_{l})_{13}$ are equal to zero in neutrino and charged lepton mass
matrices. In this case we get:

\begin{equation}
M_{\nu} =\left( \begin{array}{ccc} 0.00508&0.00381&0
\\0&0.01125&0.02987 \\0&0&0.03845
\end{array} \right)
\end{equation}

\begin{equation}
U_{\nu} =\left( \begin{array}{ccc} 0.8632&0.5047&0.01141
\\-0.39654&0.6639&0.63403 \\0.31242&-0.55183&0.77323
\end{array} \right)
\end{equation}

\begin{equation}
M_{\nu}^{D}=\left( \begin{array}{ccc}0.00455&0&0
\\0&0.00983&0 \\0&0&0.0492
\end{array} \right)
\end{equation}

\begin{equation}
V_{\nu}^{\dagger}=\left( \begin{array}{ccc} 0.96535&-0.25834&0.03693
\\0.26096&0.95498&-0.14109 \\0.00118&0.14584&0.98931
\end{array} \right)
\end{equation}

and (in $MeV$)

\begin{equation}
M_{l} =\left( \begin{array}{ccc} 0.51368&-10.7738&0
\\0&105.655&-180.361 \\0&0&1767.78
\end{array} \right)
\end{equation}

\begin{equation}
U_{l} =\left( \begin{array}{ccc}
0.99479&-0.10197&0.00004\\0.10144&0.98961&-0.10186
\\0.01035&0.10133&0.9948
\end{array} \right)
\end{equation}

\begin{equation}
M_{l}^{D}=\left( \begin{array}{ccc}0.510999&0&0
\\0&105.658&0 \\0&0&1776.99
\end{array} \right)
\end{equation}

\begin{equation}
V_{l}^{\dagger}=\left( \begin{array}{ccc} 1&0.0005 &2.99\times
10^{-6}
\\-0.0005&0.99998&0.00606 \\1.06 \times 10^{-8}&-0.00606&0.99998
\end{array} \right)
\end{equation}

\noindent we have got a solution in which mixing matrix elements
with value $0.1$ in charged lepton mixing matrix $U_{l}$ give
substantial contribution to lepton mixing matrix $U_{PMNS}$. They
are smaller then neutrino mixing matrix elements but not as tiny as
in examples of solutions given in the beginning (Eq. (24). This
solution is not unique. It is possible to get solution (with
accuracy $10^{-13}$) with higher values of neutrino masses with
slightly modified neutrino mass matrix and no change in lepton mass
matrix parameters. One should remember that in the case of neutrinos
we have one additional free parameter compared to quarks because we
do not know absolute values of neutrino masses only mass squared
differences. It is not possible to fix absolute values of neutrino
masses in this fit. We can also find a solution with smallest
neutrino mass equal to zero with $\chi^{2}\simeq 0.24$. The value of
$\chi^{2}$ is higher then for solutions mentioned before, it can not
be excluded, but is highly unsatisfactory.

With the additional universal assumption $(M)_{13}=0$ we can also
look for solutions of this type for quarks. As we know from the
previous discussion there is no problem with getting this type of
solution. We know that it is possible to give a solution for quarks
in which in addition to $(M_{u})_{13}=0$ also $(M_{u})_{23}=0$. In
this case we get:

\begin{equation}
M_{d} =\left( \begin{array}{ccc} 5.04798&13.0481&0
\\0&94.18&176.118 \\0&0&4196.3
\end{array} \right)
\end{equation}

\begin{equation}
U_{d} =\left( \begin{array}{ccc} 0.99047&0.13773&2.92 \times 10^{-6}
\\-0.13761&0.9896&0.04195 \\0.00578&-0.04155&0.99912
\end{array} \right)
\end{equation}

\begin{equation}
M_{d}^{D}=\left( \begin{array}{ccc}5&0&0
\\0&95&0 \\0&0&4200
\end{array} \right)
\end{equation}

\begin{equation}
V_{d}^{\dagger}=\left( \begin{array}{ccc} 0.99997&-0.00732&6.88
\times 10^{-6}
\\0.00732&0.99997&-0.00094 \\3.51
\times 10^{-9}&0.00094&1
\end{array} \right)
\end{equation}

and

\begin{equation}
M_{u} =\left( \begin{array}{ccc} 2.25937&-113.728&0
\\0&1244.82&0 \\0&0&172500
\end{array} \right)
\end{equation}

\begin{equation}
U_{u} =\left( \begin{array}{ccc} 0.99585&-0.09098&0
\\0.09098&0.99585&0 \\0&0&1
\end{array} \right)
\end{equation}

\begin{equation}
M_{u}^{D}=\left( \begin{array}{ccc}2.25&0
\\0&1250&0 \\0&0&172500
\end{array} \right)
\end{equation}

\begin{equation}
V_{u}^{\dagger}=\left( \begin{array}{ccc} 1&0.00016&0
\\-0.00016&1&0 \\0&0&1
\end{array} \right)
\end{equation}

We see from this solution that mixing in $(U_{CKM})_{12}$  is split,
both mixing matrices $U_{d}$ and $U_{u}$ contribute to $U_{CKM}$. It
is similar to neutrino and charged lepton case in Eqs. (36, 40).
Matrix element $(U_{PMNS})_{13}$ is reproduced by mixing in
$(M_{u})_{12}$ and $(M_{d})_{23}$ in mass matrices for up and down
quarks. Unlike in the case of leptons mass matrix for up quark is
not uniquely determined. It is possible to give examples of
solutions in which matrix element $(M_{u})_{23}$ is different from
zero ($(M_{d})_{23}$ also changes) but mixing in $(U_{u})_{12}$
(that is bigger) is not strongly influenced. We have much stronger
mass hierarchy for up and down quarks (with the mass errors
relatively much bigger then in the case of charged leptons) and
small mixing in $U_{CKM}$ in comparison with $U_{PMNS}$ so the
lepton sector is somehow more restrictive in our numerical analysis.

We also want to give an example of solution in which number of free
parameters is reduced by 3 (but still we have formally 4 free
parameters) by the relation between different mixing matrices. We
thought at the beginning that it could be an alternative to the
assumption of vanishing mass matrix elements. On the other hand from
the grand unification $SU(5)$ symmetry ($\bar{5}$ representation) we
have a hint about connection between right handed components of down
quarks and left handed components of leptons. This components being
in the same $SU(5)$ multiplet when there is a mixing between
families and in some way $SU(5)$ structure is not completely lost in
low energies \cite{lind} we would have the same mixing matrix for
right handed down quarks and left handed charged leptons. We will
now look what are the consequences of assumption $V_{d}=U_{l}$ (see
also \cite{par}). It is not clear how well this relation could be
satisfied because of corrections so we will assume artificial error
0.001 (we know that mixing in $U_{l}$ is rather weak). Minimalizing
$\chi^2$ function being sum of all terms corresponding to quarks and
leptons and taking deviations of $V_{d}$ from $U_{l}$ with
artificial error we can find solution

\begin{equation}
M_{d} =\left( \begin{array}{ccc} 4.95067&5.4794&9.94122
\\0&96.0835&84.3431 \\0&0&4199
\end{array} \right)
\end{equation}

\begin{equation}
U_{d} =\left( \begin{array}{ccc} 0.99837&0.05049&0.00237
\\-0.05709&0.99817&0.02009 \\-0.00122&-0.02020&0.99979
\end{array} \right)
\end{equation}

\begin{equation}
M_{d}^{D}=\left( \begin{array}{ccc}4.94261&0&0
\\0&96.2204&0 \\0&0&4199.86
\end{array} \right)
\end{equation}

\begin{equation}
V_{d}^{\dagger}=\left( \begin{array}{ccc} 1&-0.00294&-1.43 \times
10^{-6}
\\0.00294&1&-0.00046 \\2.79
\times 10^{-6}&0.00046&1
\end{array} \right)
\end{equation}

and

\begin{equation}
M_{u} =\left( \begin{array}{ccc} 2.24369&-214.598&993.738
\\0&1234.48&-3748.62 \\0&0&172461
\end{array} \right)
\end{equation}

\begin{equation}
U_{u} =\left( \begin{array}{ccc} 0.98522&-0.17118&0.00576
\\0.17127&0.98498&-0.02173 \\-0.00195&0.02240&0.99975
\end{array} \right)
\end{equation}

\begin{equation}
M_{u}^{D}=\left( \begin{array}{ccc}2.21053&0
\\0&1252.68&0 \\0&0&172505
\end{array} \right)
\end{equation}

\begin{equation}
V_{u}^{\dagger}=\left( \begin{array}{ccc} 1&0.00031&-2.50\times
10^{-8}
\\-0.00031&1&0.00016 \\7.49
\times 10^{-8}&-0.00016&1
\end{array} \right)
\end{equation}

\begin{equation}
M_{\nu} =\left( \begin{array}{ccc} 0.00750&0.00437&0.00333
\\0&0.01358&0.03279 \\0&0&0.03556
\end{array} \right)
\end{equation}

\begin{equation}
U_{\nu} =\left( \begin{array}{ccc} 0.82042&0.56532&0.08563
\\-0.45100&0.54778&0.70465 \\0.35145&-0.61673&0.70437
\end{array} \right)
\end{equation}

\begin{equation}
M_{\nu}^{D}=\left( \begin{array}{ccc}0.00667&0&0
\\0&0.01098&0 \\0&0&0.04945
\end{array} \right)
\end{equation}

\begin{equation}
V_{\nu}^{\dagger}=\left( \begin{array}{ccc}
0.92236&-0.38066&0.06590
\\0.38611&0.90260&-0.19034 \\0.01298&0.20101&0.97950
\end{array} \right)
\end{equation}

and in ($MeV$)

\begin{equation}
M_{l} =\left( \begin{array}{ccc} 0.51100&0.30913&-0.03127
\\0&105.658&0.81566 \\0&0&1776.98
\end{array} \right)
\end{equation}

\begin{equation}
U_{l} =\left( \begin{array}{ccc} 1&0.00293&-0.00002
\\-0.00293&1&0.00046 \\0.00002&-0.00046&1
\end{array} \right)
\end{equation}

\begin{equation}
M_{l}^{D}=\left( \begin{array}{ccc}0.510999&0&0
\\0&105.658&0 \\0&0&1776.98
\end{array} \right)
\end{equation}

\begin{equation}
V_{l}^{\dagger}=\left( \begin{array}{ccc} 1&-0.00001&6.58\times
10^{-9}
\\0.00001&1&-0.00003 \\-6.19 \times
10^{-9}&0.00003&1
\end{array} \right)
\end{equation}

We get for this solution $\chi^{2}=0.0086$ with formally 4 free
parameters. This solution is not unexpected. Small mixing in $U_{l}$
enforces by the condition $V_{d}=U_{l}$ small mixing in $V_{d}$ and
it means that mixing in $U_{CKM}$  comes mainly  from the mass
matrix for up quarks. The obtained solution corresponds to the
solution for $M_{u}$ and $M_{\nu}$ matrices given in Eq. (32) and
Eq. (33). The mass matrices given in Eq. (32) and Eq. (33) with
diagonal matrices for down quarks and leptons satisfy $SU(5)$
condition as unit matrices. It is difficult to find a solution with
a small $\chi^2$ value corresponding to the situation where most of
the mixing in $U_{CKM}$ comes from down quark and only small part
from $M_{u}$ mass matrix. It is not possible to find such solution
with additional assumption that in all mass matrices matrix element
$(M)_{13}$ vanishes.

We have tried to determine mass matrices for quarks and leptons
assuming that they have triangular form. For all particles it is
assumed that mass matrices are upper triangular matrices. It means
that in the decomposition of given mass matrix into diagonal and two
unitary matrices left handed components of fields determine the
mixing matrices. For hierarchical mass spectrum  mixing in left
handed components is much stronger then in the right handed
components. As an input we use the values of matrix elements for CKM
and PMNS matrices, masses of quarks and charged leptons with known
errors and neutrinos squared mass differences with corresponding
errors. With these assumptions one can not determine the elements of
mass matrix uniquely using fit to experimental data. We have
discussed examples of solutions for quarks and leptons with
relatively small non diagonal matrix elements. It was stressed that
that the masses of charged leptons are known with extremely high
accuracy (the error in the determination of the mass of electron is
comparable with naively expected highest neutrino mass). Together
with very strong mixing in lepton sector, lepton sector seems to be
more restrictive then quark sector. Considering various numerical
solutions for leptons we came to the conclusion that in order to
restrict the number of possible solutions and treat neutrinos and
charged leptons in the same way we have to assume dynamical
condition $(M_{\nu})_{13}=0$ and $(M_{l})_{13}=0$. We have  given
the solutions under these assumptions. Unfortunately, that does not
fix the absolute scale for neutrino masses. Then we have extended
this assumption to up and down quarks. The experimental
justification for this assumption is that in $U_{CKM}$ and
$U_{PMNS}$ mixing matrices matrix element $(U)_{13}$ is much smaller
then the other mixing matrix elements. We have also at the beginning
as alternative to vanishing mass matrix elements considered other
possibilities. With the additional assumption which connects mixing
matrices for right handed down quarks and left handed charged
leptons in the way suggested by $SU(5)$ symmetry it is possible to
find a solution for mass matrices with weak mixing for down quarks
and charged leptons. The mixing is stronger for neutrinos and up
quarks (mixing of left handed up quarks is mainly responsible for
mixing in CKM matrix). With our additional assumption for all mass
matrices $(M)_{13}=0$ we can not find such solution.

\end{document}